# Microstructure and Aging Dynamics of Fluorescent Nanoclay Co-gels of Laponite and Montmorillonite

**Ravi Kumar Pujala[1,2]* , and H. B. Bohidar[2,3]**

[1] Soft Condensed Matter group, Debye Institute for Nanomaterials Science, Utrecht University, Princetonplein 5, 3584 CC Utrecht, The Netherlands

[2]School of Physical Sciences, Jawaharlal Nehru University

[3]Special Centre for Nanosciences, Jawaharlal Nehru University

New Delhi-110067

**Abstract**

The phenomenon of aging in soft matter systems is an intriguing and interesting. In his study we investigate the microstructure and aging dynamics of fluorescent nanoclay gels formed from the mixture of small (~25) and large (~250) aspect ratio nanodiscs using various experimental techniques. We examine how the presence of one kind of clay in other clay of different nature affect the aging dynamics, microstructure and the final resulting phase. Viscosity of the dispersions grows as a power-law with waiting time of the system and the dispersions initially show Maxwellian behaviour but deviated as the time progressed. Fractal dimension, $d_f$ of the gels is found to be 2.5. Confocal micrographs revealed clear grain boundaries in the nanoclay gels, since Na-Montmorillonite exhibit fluorescence. We also report for the first time that the gels containing sodium MMT gives a bright green-yellow fluorescence in ultra violet light. These fluorescent gels may be useful in industry and academic research.

**Introduction**

Colloidal systems mimic the atomic system due to their ability to control interaction strengths at many length scales. Due to the presence of anisotropic shape combined with the inhomogeneous distribution of many such phase states are possible [1-4]. Nanoclays are ubiquitous in nature. It has rich in dynamics and used in industries also it's the interest of scientist in making nano materials, nano composites and biocompatible products. Mostly clay as used as rheological modifiers in cosmetics, pains and food products. They earned their own place in daily life. The phenomenon of *aging* is still poorly understood and still under investigation. Aging is defined as the changing the physical properties of the system with waiting time. The properties of the system keep changing as the time progressed and the system moves to different phase states exploring the minimum possible state. In the process it may end up with gels or glasses depending on the concentration of solid particles and the interaction strength. Both the gel and glass are non equilibrium states trying to achieve its minimum energy configuration. The situation is glasses is rather even more interesting, in general the system is trapped in a state where all the particles trying to reach a minimum energy configuration. Particles exhibit two kinds of motions, in cage and out of cage motions [Springer Thesis 2014].

The origin of fluorescence from clay suspensions is still unclear since all the clays do not exhibit the fluorescence emission. Sodium Montmorillonite is one of the clays which exhibit significant fluorescence yield, other like Kaolinite, Illite do not exhibit the same. But it is suspected that the emission comes from the exchange of sodium ions [5, 6]. Here in this study we take the advantage of fluorescent clay in presence of non fluorescent clay to study the gel structure using confocal imaging and of course this is the one of the first studies of such kind.

Recently we discovered the experimental existence of equilibrium gels and empty liquids in aspect ratio Na-Montmorillonite [6]. We also made observation of gels formed through different routes: one through phase separation and another through equilibrium gelation for an extensively aged system of Montmorillonite. Existence of equilibrium gels and glassy phases also observed in Laponite nanoclays experimentally [7, 8] and also in patchy colloids [9, 10]. Ordering or the Laponite particles at water-air interface [11] and the aggregation behaviour of nanoclays in hydrophobic environment namely alcohol solutions [12] and ionic liquids have also investigated in great detail [12a].

Recently we investigated the dynamics of one clay in presence of other clay different nature and produced variety of phase states namely gels, repulsive glasses and ordered glasses with interesting aging dynamics [13-15]. There are some interesting theory and simulations works by Lekkerkerker et al and interesting review on the mixed hard platelets where they observed multiphase coexistence [16, 17]. Thus the mixed suspensions of large and small platelets give rise to numerous phases and rich in dynamics and there exist the competing interactions in arrested states of colloidal clays [18].

Here in this study we explore the microstructure and aging dynamics in the mixed clay system of two nanoclays of different aspect ratio for various compositions. Presence of larger particle in smaller particles or vice versa will change the dynamics more significantly and the new system is different from single component systems. The effect of aging and temperature in these dispersions was also examined. The dynamics were captured through light scattering, rheology measurements. All the dispersions we have studied exhibit aging and spontaneously reached an arrested state. No phase separation was observed even after months unlike single component systems which undergoes phase separation into colloid rich and colloid poor regions [6-8].

## Experimental section

### Material and Methods

Laponite RD, Nano-clay and Cloisite-Na (MMT) were purchased from Southern Clay Products, USA and used without further modification. Fractionation of MMT was done as described in our previous work since the native clay has many fractions ranging from few nanometres to micrometer. The fluorescent yielding for the native clay is lower than the fractionated clays. Different concentration ( 0.50-2.2 % w/v of Laponite (L) and MMT (MMT or M) dispersion were prepared by dispersing Laponite and MMT separately in desired amount in deionized water at room temperature and mixing them vigorously for 2 hr using magnetic stirrer. Complex of nano-clays for different mixing ratio (L: M = 1:0.5, 1:1 and 1:2) were prepared by mixing 0.50-2.2 % w/v of L and M respectively in different proportions to generate optically clear binary mixtures. The suspensions pH was adjusted to 9.0.

Rheological measurements were performed on controlled stress AR-500 Rheometer (TA Instrument, England) using small oscillatory shear in the oscillatory mode. Stainless steel plate geometry (radius 25 mm) was used during the experiments. Linear and nonlinear rheology tests were performed with different modes. Measurement of viscosity was done using a Sine-wave Vibro Viscometer (model – SV10, A and D company, Japan).

Dynamic Light Scattering (DLS) and Depolarized light scattering (DDLS) experiments were performed on a 256 channel digital correlator, (PhotoCor Instruments, USA) that was operated in the multi-tau mode (logarithmically spaced channels). The time scale spanned 8-decades, i.e. from 0.5μs to 10s. This instrument used a 35mW linearly polarized He:Ne laser . A Glan Thomson analyzer was used in front of the photomultiplier tube to enable collection of polarized and depolarized components of scattered light ($I_{VV}$ and $I_{VH}$). The probe length scale is defined by the inverse of the modulus of the scattering wave vector q where the wave vector $q = (4\pi n/\lambda)\ sin(\theta/2)$, the medium refractive index is n, excitation wavelength is $\lambda$ (=632.8nm) and $\theta$ is scattering angle. The scattered intensity values were measured at the scattering angle $90^0$. Further details on depolarized light scattering can be obtained from ref. [32].

Laser confocal microscope was used for recording optical images (Olympus Fluo View, Model FV1000). During measurements samples were placed between the microscope slide and cover slip. The steady state fluorescence spectrum was recorded in DCM using a Varian Cary Eclipse Fluorescence Spectrometer. Particle morphology was examined by JOEL 2100F transmission electron microscope (Digital TEM with image analysis system and maximum magnification = × 15,000,000) operating at a voltage of 200 KV. The aqueous dispersions were drop-cast onto a carbon coated copper grid which was air dried at room temperature (20 $^oC$) before loading onto the microscope. SEM images were captured using Zeiss EVO40. The morphology of the sample were studied from the SEM micrographs.

### Visual Observation

Initially as shown in Fig.1 after mixing the two clays dispersion in different mixing ratio complex sample remain in the sol form but eventually with waiting time the interaction between Laponite and MMT becomes coarsened in all direction i.e. the composite of M-L-M or L-M-L starts interacting also to gives rise to a 3-D network. Solution state and gel state properties were studied in systematic way in our previous study. The complexes show the intermediate properties between the individual systems. As the sample ages the physical properties keep changing and try to move towards the equilibrium or a more stable state. To understand aging behaviour we followed the dynamics using rheology, light scattering and dilution experiments.

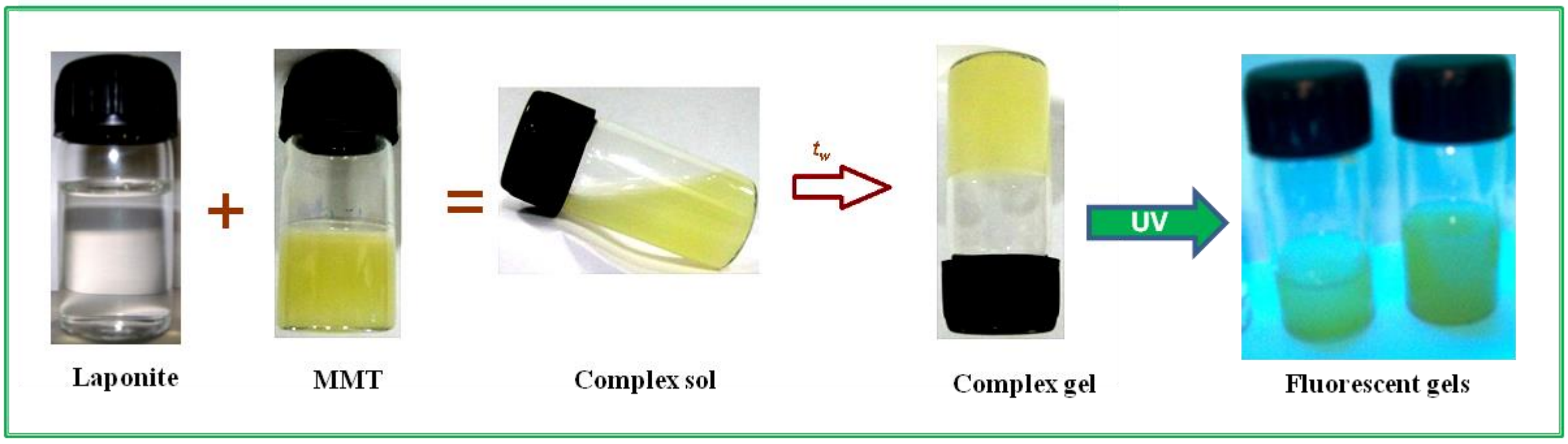

**Fig. 1** Sample preparation and evolution of different phase states (sol and co-gel) in the mixed system of Laponite and Montmorillonite.

## Confocal Imaging

Nanoclay platelets show fluorescence behaviour in aqueous medium. Fluorescence emission spectrum from clay dispersions is shown in Fig.2 for different compositions of clays. Significant fluorescent emission was seen in pure MMT than in the mixtures and Laponite shows the least or no emission. Only MMT has the capability of fluorescent yielding in water. Other clay systems like Kaolinite and Illite also haven't shown significant emission. We exploited this property from MMT and used in the imaging. This is the first experimental observation ever made on MMT clays. Figure 3 shows fluorescence behaviour of the MMT particles or aggregates of the same both at higher (upper panel) and lower (lower panel) concentrations of MMT.

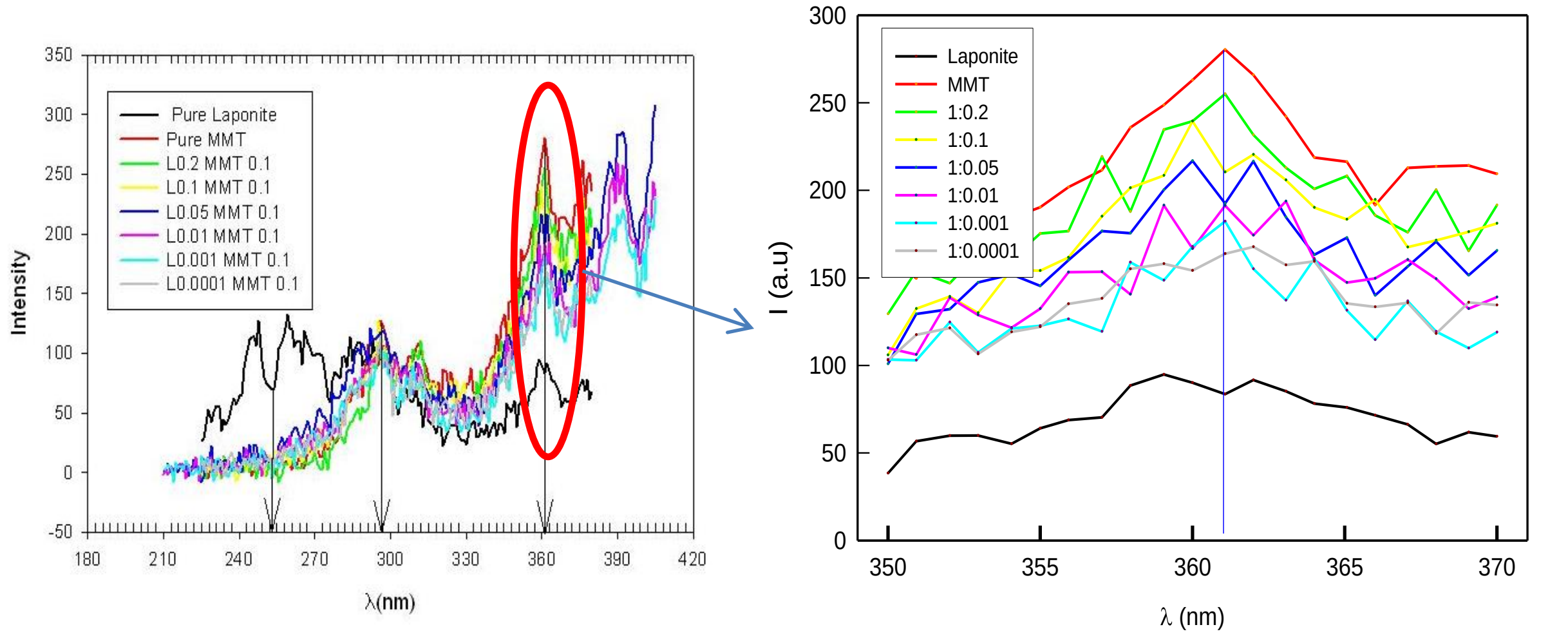

**Fig. 2** Fluorescent emission peaks of the dispersion for difference compositions as indicated.

**Viscoelastic studies**

have complete region of aging. Growth of viscosity as function of waiting time is shown in the Fig. 1. Growth of the viscosity of the cogel is in between the single component dispersions of

Waiting time $t_w$ is defined as the time elapsed between the sample preparation and measurement. Viscosity of the dispersions has been monitored for long time to Laponite and MMT. The relative viscosity $\eta_r$ grows as a power-law for the dispersions at all mixing ratios

$$\eta_r \sim t_w^{\alpha} \qquad (1)$$

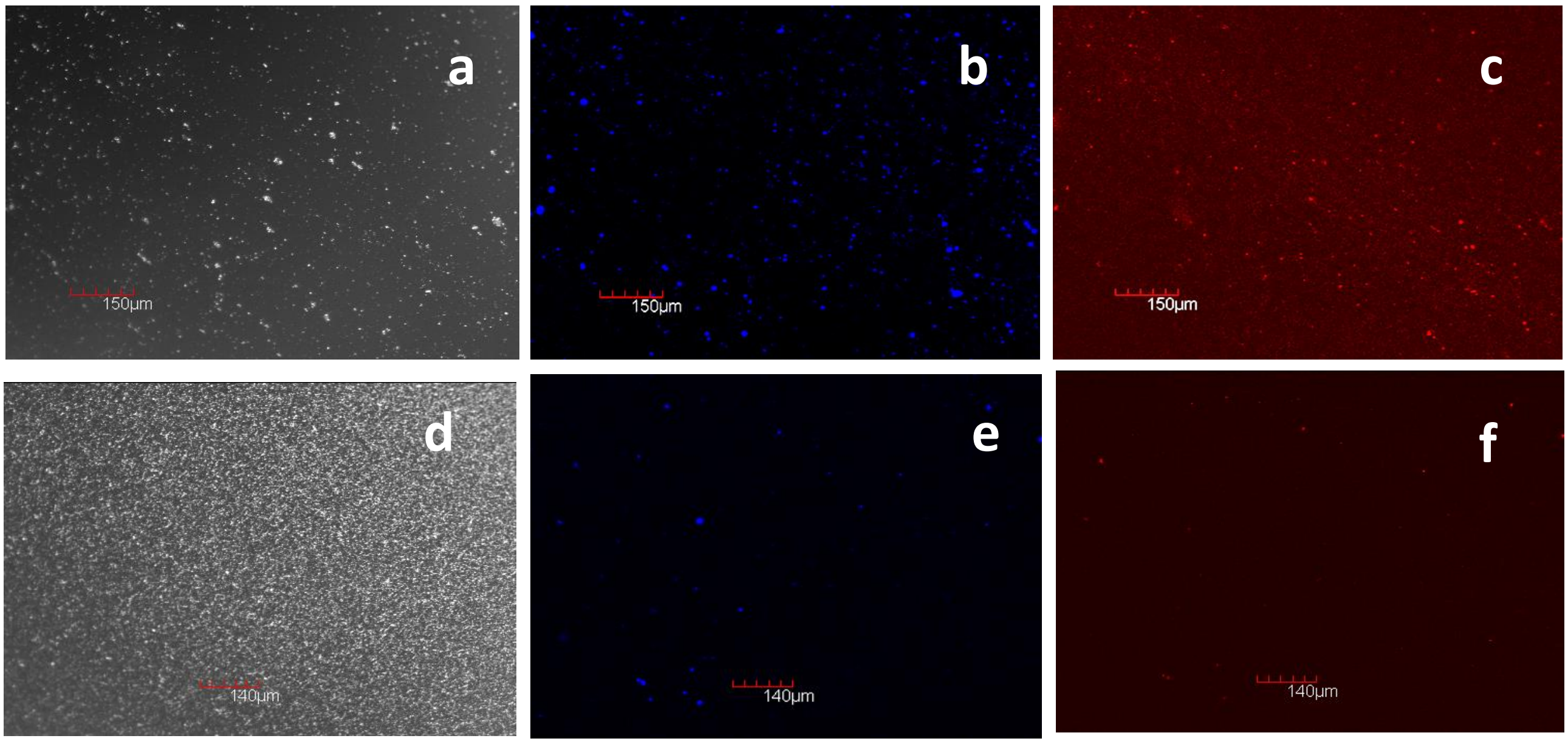

**Fig. 3** Optical and confocal Images of for higher MMT particles (top panel) and lower MMT particles (bottom panel) under different lights show the fluorescence of Montmorillonite particles in solution phase.

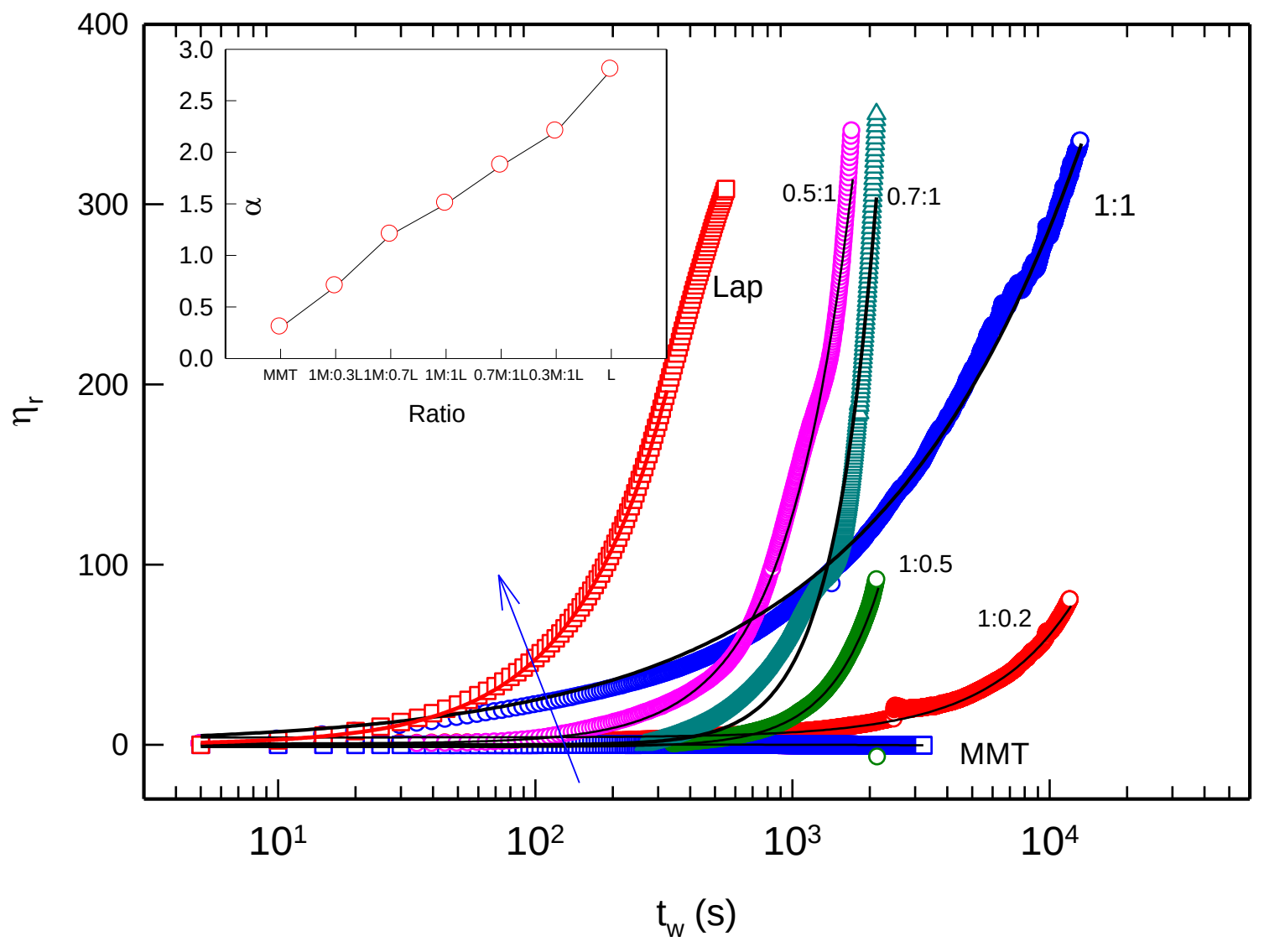

**Fig. 4** Evolution of relative viscosity as function of waiting time for various compositions. Values of the power-law exponents described in Eq.1 is shown in the inset.

The values of exponent $\alpha$ increases with mixing ratio as shown in Fig. 1 inset. The relative viscosity of the clay dispersions of small aspect ratio is always higher than the particles with high aspect ratio for the same volume fraction. Viscosity of the dispersion is affected by the particle size, shape and temperature. Smaller particles have the high accessible space compared to the larger particles for the same volume fraction. In the mixtures the excluded volumes effects comes into picture as explained in the previous work [Langmuir]. As the time progressed the particles starts interacting and lost individual nature and form a percolating network or trapped in the crowded environment such as glass state. Figure 1 indicated the smaller particle system enters the non-ergodic phase rather quickly than the larger ones. It is also noted that the smaller particles.

We followed the similar protocol of rejuvenation applied in clay systems. The sample was first loaded on to the Rheometer and used the solvent trap for long time aging experiments. The sample is first shear melted at the rate 800 $s^{-1}$ before any measurement. Frequency sweeps were performed at different waiting times as shown in Fig.3. It was found that the system showed a Maxwellian behaviour in initial aging period. Aging affects the dispersions and move from an ergodic to non-ergodic state. During initial aging period the system follows the viscoelastic behaviour defined by Maxwell.

Time-dependent rheology studies were performed on the dispersions in the frequency sweep mode to probe the visco-elastic properties of the samples. According to Maxwell model, one can express the in-phase modulus *G'* and out of phase modulus *G"* as function of frequency as

$$G' = \frac{G_0 \omega^2 \tau^2}{1+\omega^2\tau^2} \quad \text{and} \quad G" = \frac{G_0 \omega \tau}{1+\omega^2\tau^2} \tag{2}$$

where $G_0$ is the plateau elastic modulus, $\tau$ is the relaxation time, and $\omega$ is the angular frequency. In the limiting case $\omega \to 0$, the relations will become $G' \sim \omega^2$ and $G'' \sim \omega$ which are considered signatures of the Maxwellian model. Deviation from the Maxwell behaviour was observed and the system was probed systematically. The power, $n$ starts decreasing with waiting time from 2 and 1 for *G'* and *G''* respectively in an ordered manner and reaches a plateau where there is no change in slope observed at long waiting times and system reaches fully arrested state. It is plotted in Fig. 4 with waiting time. An interesting evolution of slopes was found in the aging system. When the system is fully aged it becomes solid and the frequency dependencies vanished. The same is observed for all the compositions that were probed in this study.

The frequency dependent modulii at different waiting times render additional information on the evolution of viscoelasticity during gelation process as shown in Fig. 5. For short times *G'* and *G"* display marked frequency dependence with *G">G'*. and at the long time behaviour is observed as solid-like response with *G'>G"*. The more interesting phenomenon observed was at time $t_w \sim t_g$, $G'$, $G" \sim \omega^n$. Gelation time $t_g$ observed is close to the critical time observed in viscosity measurements. As shown in Fig. 6 there is no crossover of G' and G" below $t_g$ which indicated liquid like behaviour below $t_g$. When $n$ reaches 0.5 the magnitudes and the frequency dependence of G' and G" become equal and beyond which *G'>G"* and have the solid-like response.

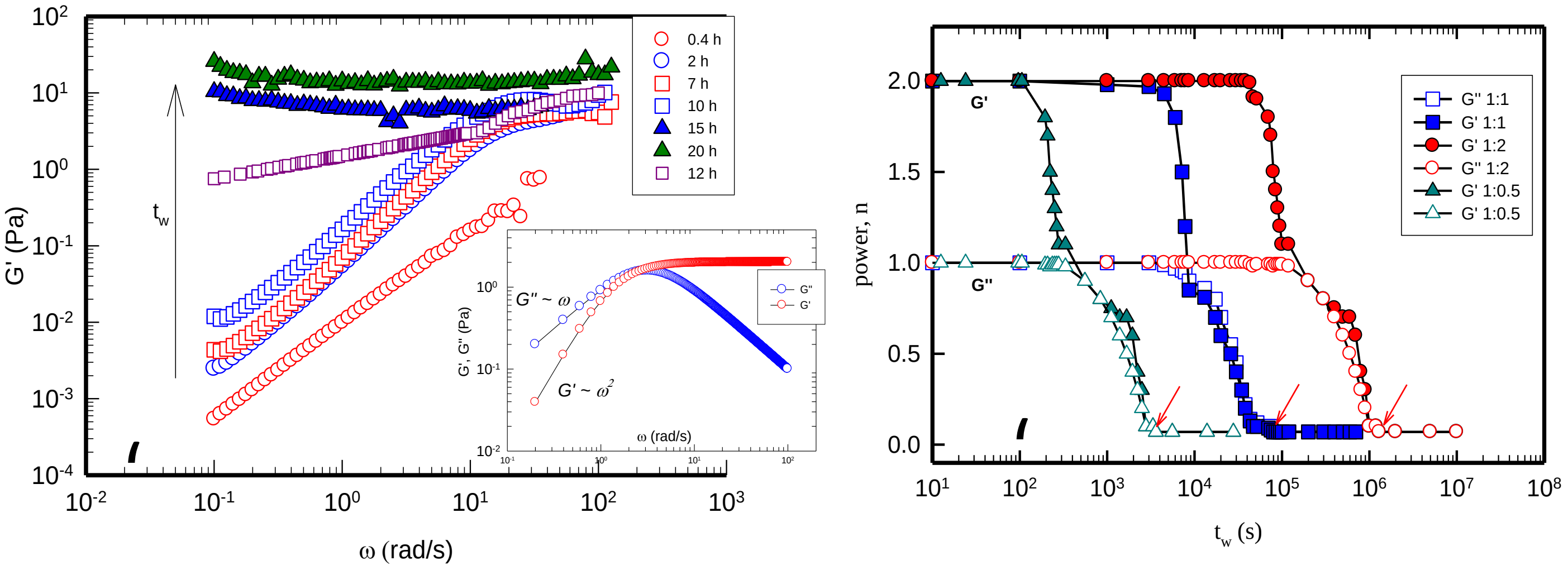

**Fig. 5 A)** Dynamic frequency sweeps at different aging times. The storage modulus reached a plateau at long waiting times. Viscous modulus is not show to make the plot clear. Equal proportions of clays (1:1) are used for this measurement. Arrow indicates the direction of growth of modulus. Inset shows the Maxwellian behaviour for the freshly prepared samples. **B)** Evolution of the power *n* (the slope of the *G* vs *ω* curves at low frequencies as described above samples) as a function of waiting time.

This indicates that the sample although it's not a chemical gel, the interparticle interactions are so strong that the relaxation time of the system takes place beyond the experimental window. The power-law scaling of modulus with frequency suggests self-similar relaxation which is expected for the critical gel. The cross over time fairly provides the accurate measure of the gelation time since effectively independent of frequency as G′ and G″ are scaled by the same factor, $cos(n_c\pi/2) = sin(n_c\pi/2) = 1/\sqrt{2}$ for $n_c = 0.5$. A close observation at $G' \approx G'' \sim \omega^{0.5}$ suggests that the critical gels not only have the same frequency dependent viscoelasticity, but that the magnitudes of the moduli are similar. It is evident from the above picture that the system initially a liquid passes through a critical state and become solid beyond the critical point or the gelation time.

For a gelation due to strong attractive interactions ($E/k_BT$>>1) ($E$ is a depth of attractive potential) sol state transform into a gel one at a well-defined gelation point, $t_{gel}$ ( $t$ is a time) At $t_{gel}$ system exhibit a behaviour specific to percolation such as a power law frequency dependence of complex modulus which reflect the percolation transition. Dynamic viscoelastic modulii exhibit a power law frequency dependence at gelation point $t_{gel}$,

$$G' \approx G''(\omega) \approx \omega^u \tag{3}$$

Where *G'* and *G''* are real and imaginary part of complex modulus and $\omega$ is angular frequency. Here $u$ is related to the fractal dimension $d_f$ of gel network structure and spatial dimension $d$ as (In the case of unscreened excluded-volume interactions percolation theory predicts) $u = d/(d_f+2)$, where $d_f$ is the fractal dimension of the polymer. If we assumed instead screened excluded-volume interactions we would expect $u = [d(d+2- d_f)]/[2(d+2- d_f)] = 0.5$ for $d_f = 2$, which is exactly similar to our experimental results. At this gelation point an infinite percolated network is formed and thus static viscosity $\eta$ diverges [Fig. 4].

## Dynamic Light scattering

Aging dynamics of dispersions were best probed using dynamic light scattering techniques by evaluating the dynamic structure factor with waiting time by taking out the heterodyne contribution as described in our previous work [--]. In general the dynamics structure factor in clay systems is described by a two step relaxation [---],

$$g_1(q,t) = a\exp-\left(\frac{t}{\tau_1}\right) + (1-a)\exp-\left(\frac{t}{\tau_2}\right)^{\beta} \tag{4}$$

where $a$ and $(1-a)$ are the weights of the two contributions $\tau_1$ and $\tau_2$, which in turn are the fast and the slow mode relaxation times respectively, and $\beta$ is the stretching parameter. The fast mode relaxation time $\tau_1$ is related to the inverse of the short-time diffusion coefficient $D_s$ as $\tau_1=1/D_sq^2$.

For times less than the ergodicity breaking time, the fast and slow relaxation times varies inversely with the square of wave vector ($\tau_1$, $\tau_2 \sim q^{-2}$), a feature reminiscent of classical diffusion. Beyond the ergodicity breaking time the slow relaxation time varies inversely with the wave vector ($\tau_2 \sim q^{-1}$) which show the diffusion other than the classical diffusion.

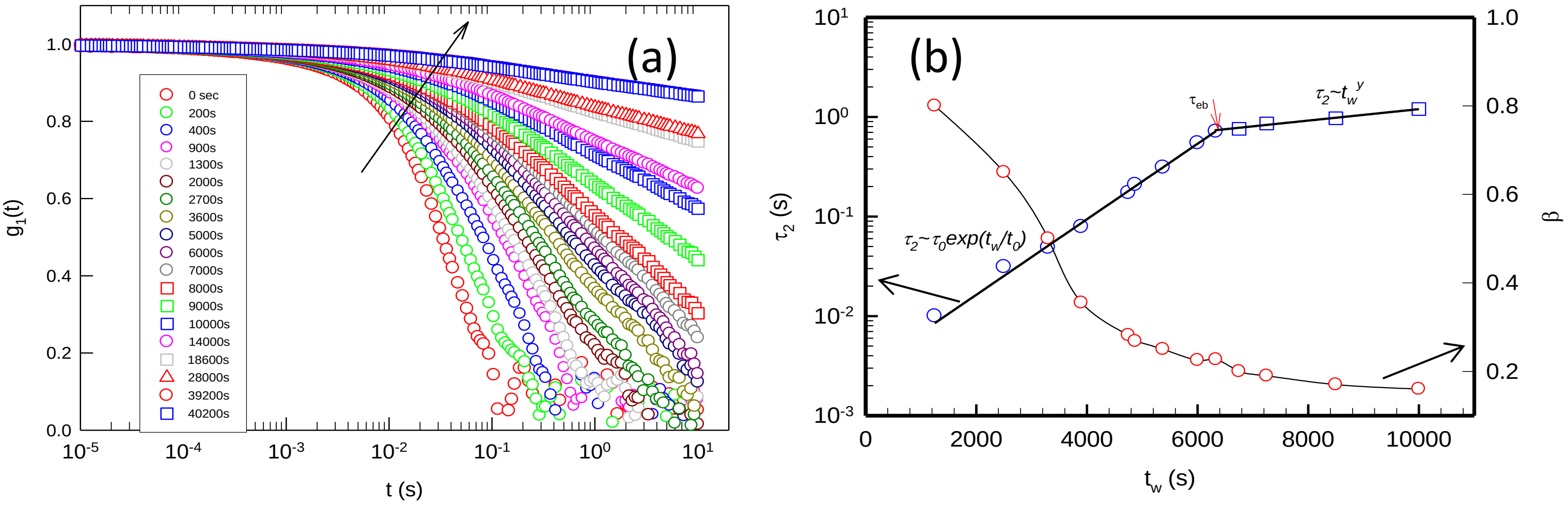

**Fig. 6 a)** Evolution of dynamic structure factor as function of waiting time. Arrow indicates the direction of increment in waiting time. **b)** Characteristic slow relaxation time and the stretch exponent β as function of $t_w$.. Note the two regions are differentiated by ergodicity breaking time.

The short time diffusion motion corresponds to the classical diffusion and shows no change with aging. The characteristic fast relation time remained invariant throughout the aging process. Characteristic slow relaxation time grows exponentially with waiting time $t_w$. The temporal evolution of slow mode relaxation time $\tau_2$ is presented in Fig. 8 for $t_w<\tau_{eb}$.

$$\tau_2 \sim \tau_0 \exp(t_w / t_0) \tag{5}$$

The temporal evolution of slow mode relaxation time $\tau_2$ in full aging regime i.e. above $\tau_{eb}$ $(t_w<\tau_{eb})$, follows a linear growth as

$$\tau_2 \sim t_w^{\ y} \tag{6}$$

Fig.7 shows the optical images of the cogels for three mixing ratios (1:1, 1:0.5, and 1:2). Fig 5B correspond to equal amount of Laponite and MMT, it shows the formation of homogeneous domains in the complex system whereas fig 7C and 5A correspond to more MMT and more Laponite respectively in the complex system. These figures shows that system with more MMT (less Laponite) has less connectivity hence less number of domains are present in the system whereas more Laponite in the system make the system very dense and homogeneous.

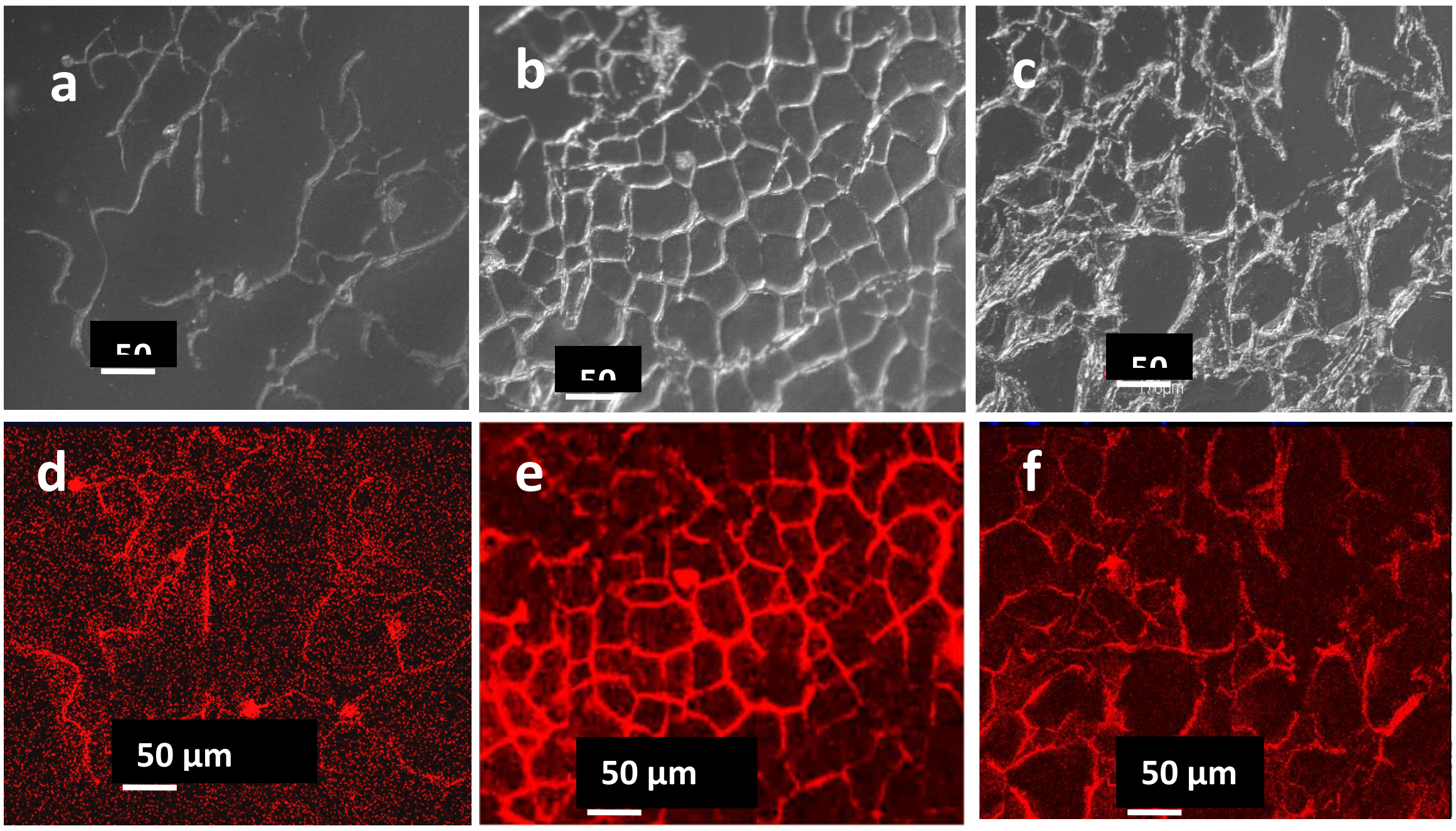

**Figure 7**: Optical (top panel) and the corresponding confocal micrographs (bottom panel) of the grain boundary network of complex gel for 1:05, 1:1 and 1:2.

In order to probe the structure more closely we obtained the SEM and TEM micrographs shown in Fig. 10. A close look at the micrographs gives information of the morphology of the mixed clay gels. Homogenous structures are seen in the gels with equal mixing ratio (Figs. 10B and 10E). Strand like morphology is clear from the cogels of equal compositions which is similar to our previous observation on MMT gels.

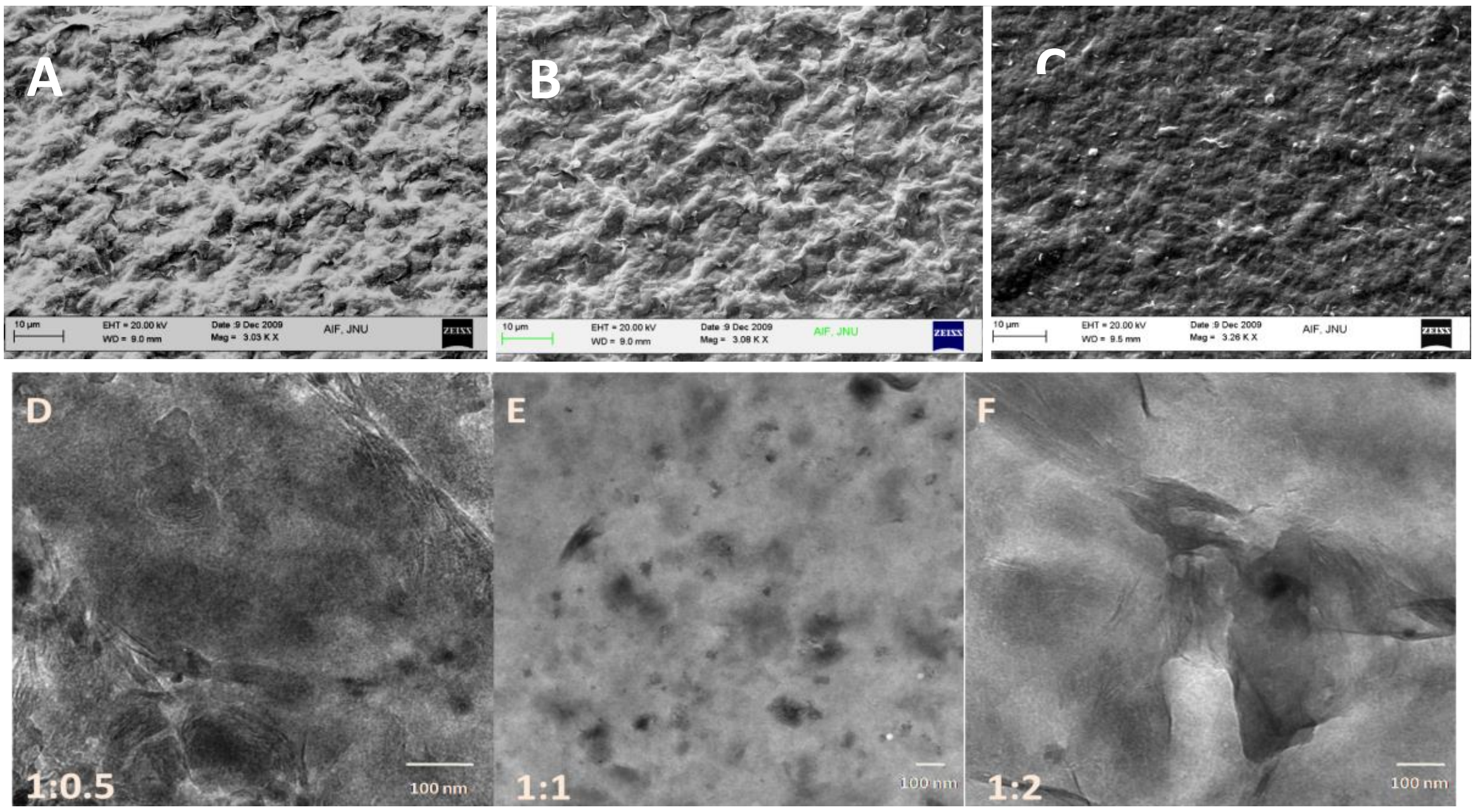

**Figure 8**: SEM micrographs (top panel) and the corresponding TEM images (bottom panel) co-gel for 1:05, 1:1 and 1:2.

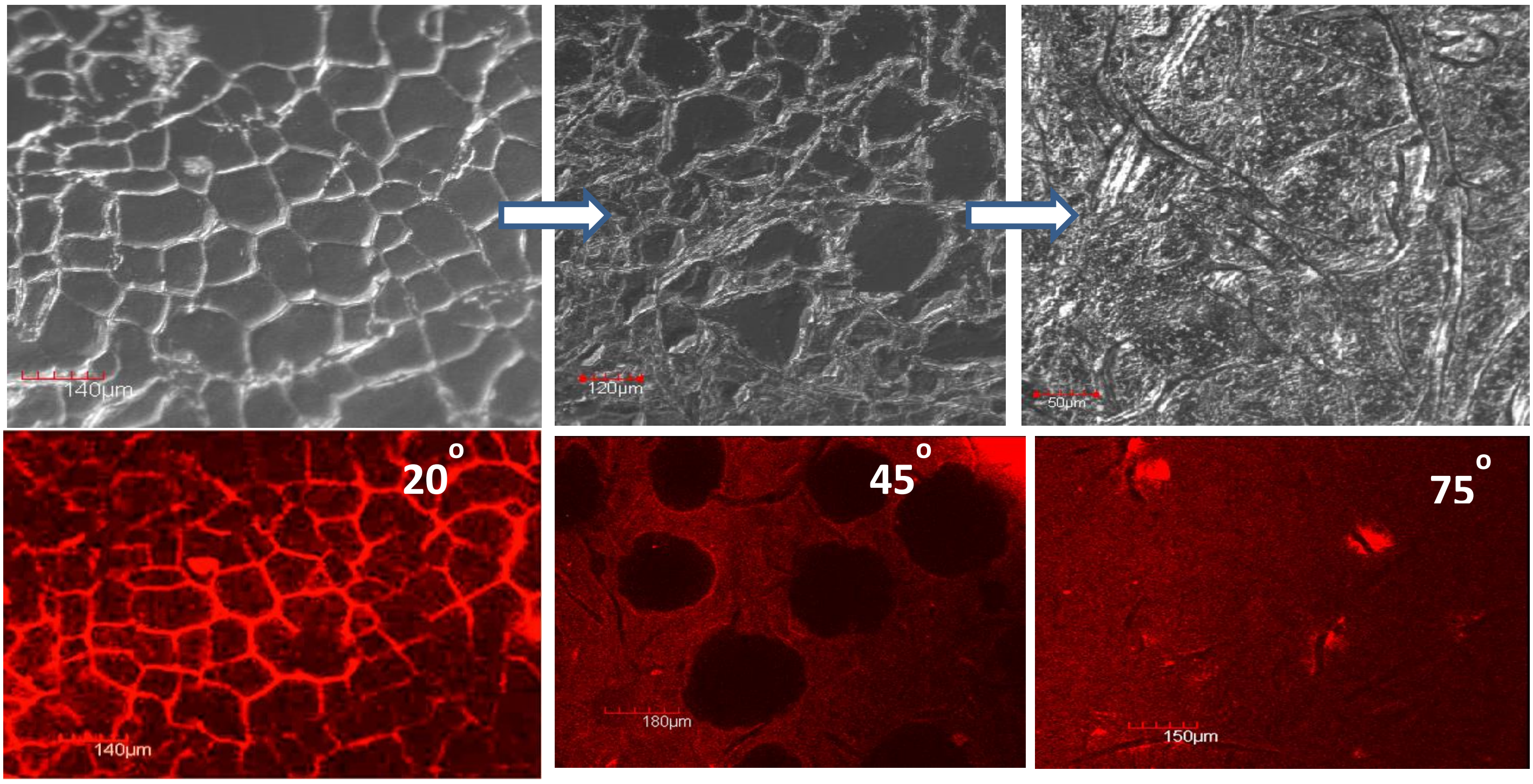

**Fig. 9** Optical micrographs of gel (1:1) at different temperatures indicating the irreversible thermal transition.

**Temperature Induced Irreversibility**

Gels made of clays are physical in nature and exhibit strong viscoelastic behaviour similar to chemical gels. Theses gels found to age irreversibly with waiting time and temperature. As the gel is heated the arrangement of larger particles gets randomly distributed and becomes more heterogeneous and still retain the gel structure with larger elastic modulus. The domains in the system get disrupted with temperature making the larger particles to get out of the network and enter a glassy state which was observed from the dilution experiments and also from the micrographs of the samples in the in the glassy state and the viscoelastic length decreases (Fig S4 Supporting Information). The glassy state also confirmed from the dilution experiments. Dilution tests are done by adding water to the sample which was quenched, the solvent was able to melt the existing state confirms the domination of repulsive interactions.

**Discussion**

Here in this study we observe for the first time the fluorescent behaviour of clay cogels without adding any fluorescent pigments to the natural clay. Small amounts of the synthetic Laponite particles produce gel structures in smectic clays (Na-MMT). Small platelets are able to induce the gel structure in an otherwise viscous system of large particles. Small platelets help the bigger particles bind together through electrostatic attractions and the structure grows with aging, coarsened in all directions giving rise a network structure and increases the viscosity and also elastic modulus. Viscosity of the mixed suspension is always higher than the simple system which is still to be understood, it may due do the directed interactions. As the concentration of the particles increases, the anisotropic interactions between particles of different nature make them form either T bonds or band structure as discussed in our previous studies [6], as a result the viscosity, elasticity and the strength of the system increase and behaves as solid.

Aging dynamics in mixed clay system is similar to the single component system where it exhibits more than one relation times. During initial stage of aging, relaxation time grows as fast as exponential followed by a linear growth. Similar picture is evident from rheological measurements, during initial aging it behaves as Maxwell fluid, but as time progresses it gets deviated from the Maxwell behaviour and crosses to a solid like behaviour through a critical gel state. Beyond the ergodicity breaking time or the cross over time solid nature is apparent. The fractal dimension of the mixed system is found to be 2.5, comes from the screened excluded volume effects. Thus the gel is made of particles of different nature with anisotropic interactions.

The fluorescence yielding of the cogels varies according the composition of the individual clay particles and is mostly influenced by volume fraction of large particles. A clearer picture emerges from the confocal measurements where the bigger platelets form the network inside the dispersions made of small platelets. The strength of network formed by the MMT particles heavily depends on the mixing ratio, amount of MMT and the nature of interactions between them. Even in the mixed suspensions the nature of bigger particles is not lost and they interact in a specific way to build the network of their own for any mixing ratio. This same picture emerged from the Fig.7 (a-c). The regions we found here is similar to the grain boundary network of a colloidal polycrystal, where fluorescent particles sit the grain boundaries of the polycrystal [19]. Here in our case large MMT particles sit at the specific locations in the dispersion of small platelets forming network of their own. For the equal mixing ratio the dispersion found to be more homogeneous and is in accordance with our previous observations. Similar picture is also emerged from SEM and TEM measurements. The network is more homogeneous for equal mixing ratios. High percent of small platelets make the network appear to smoothen the network observed from Fig. 8. This study makes us believe that the dispersion is more homogeneous with equal mixing ratio though the number of small and large particles for the give volume is different.

To observe the effect of temperature 1:1 complex was chosen. Confocal micrographs show the rupture of network formed by the MMT particles and a thermal irreversible transition takes place with increasing elasticity with a clear dehydration transition temperature as shown in Fig. S4. Viscosity of the thermally heated gels is irreversible and shown to increase during cooling (Fig. S3). Thus the temperature will continue to play a significant role in soft matter systems like clays to induce irreversible thermal transitions and increase viscoelastic nature of the system.

**Conclusion**

To conclude, we report here for the time in the literature that the gels containing sodium MMT gives a bright green-yellow fluorescence in ultra violet light, where as pure Laponite gels don't give any fluorescence. We studied the aging dynamics and microstructure of these fluorescent gels formed from the clays of different aspect ratios (250 and 25) and different. Initially mixed clay dispersions show Maxwell viscoelastic nature and they go through a critical gel state and deviate from Maxwell behaviour. There present two kinds of relaxations in the system namely fast and slow. During initial aging the slow relaxation time increases exponentially and in the later aging period it grows linearly similar to other gel and glassy systems. Microscopic studies allowed us to understand the microscopic networks formed by the clays, where the larger platelets form the network of their own in presence of smaller particles. Thus the Montmorillonite clays could be used as fluorescent probes in imaging though it shows lower lever fluorescence compared to the strong fluorescent materials. We also found a thermally irreversible transition in mixed clay gels. Thus our studies on mixed clay system will continue to attract and guide the scientist for a long to come though better understanding of the interactions is needed.

**Acknowledgments**

RKP acknowledges receipt of Junior and Senior Research Fellowship from Council of Scientific and Industrial Research, Government of India.

**Supporting Information**

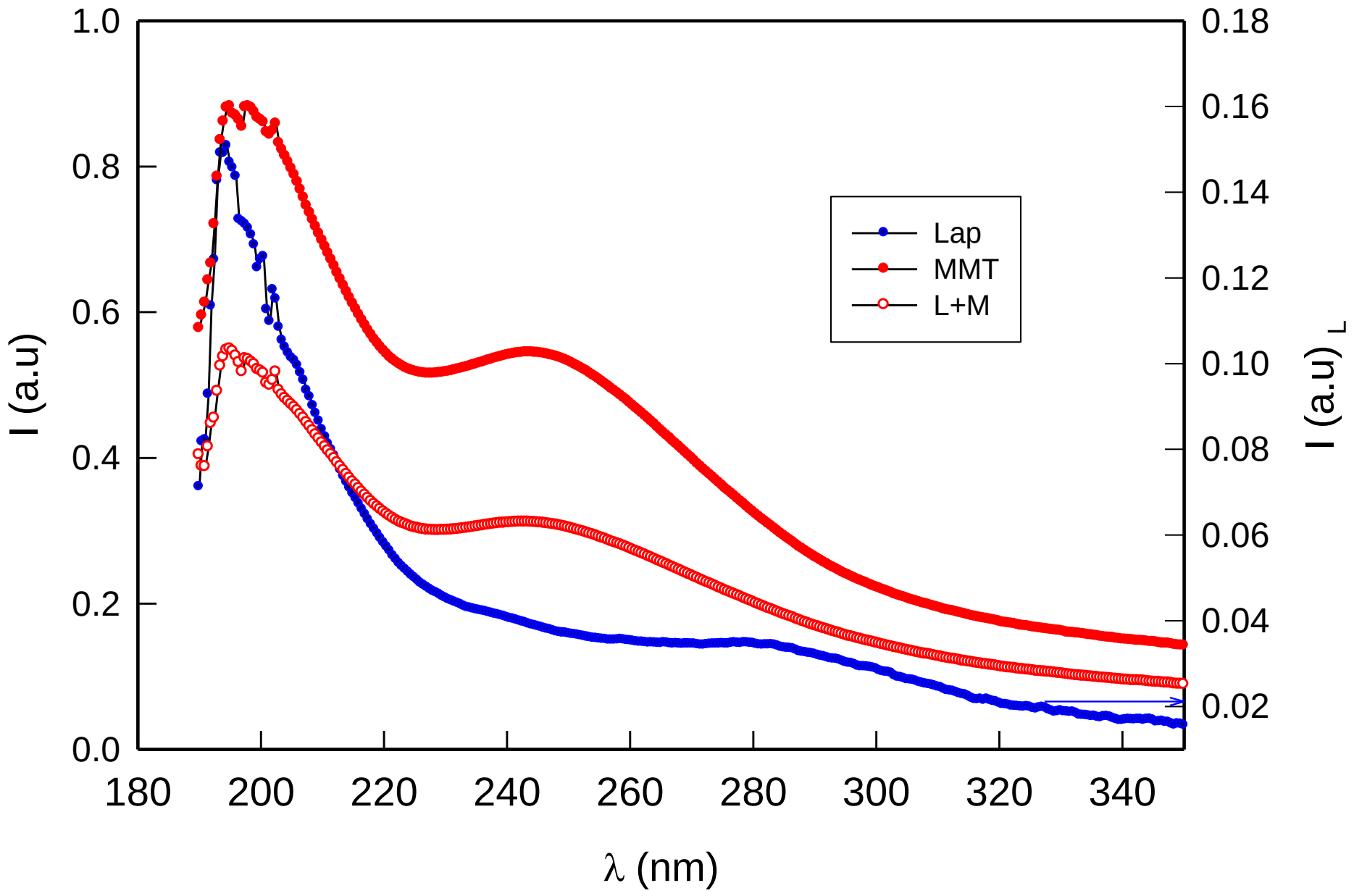

**Fig. S1** UV-absorption spectra of Laponite, Montmorillonite and its composite in equal mixing ratio

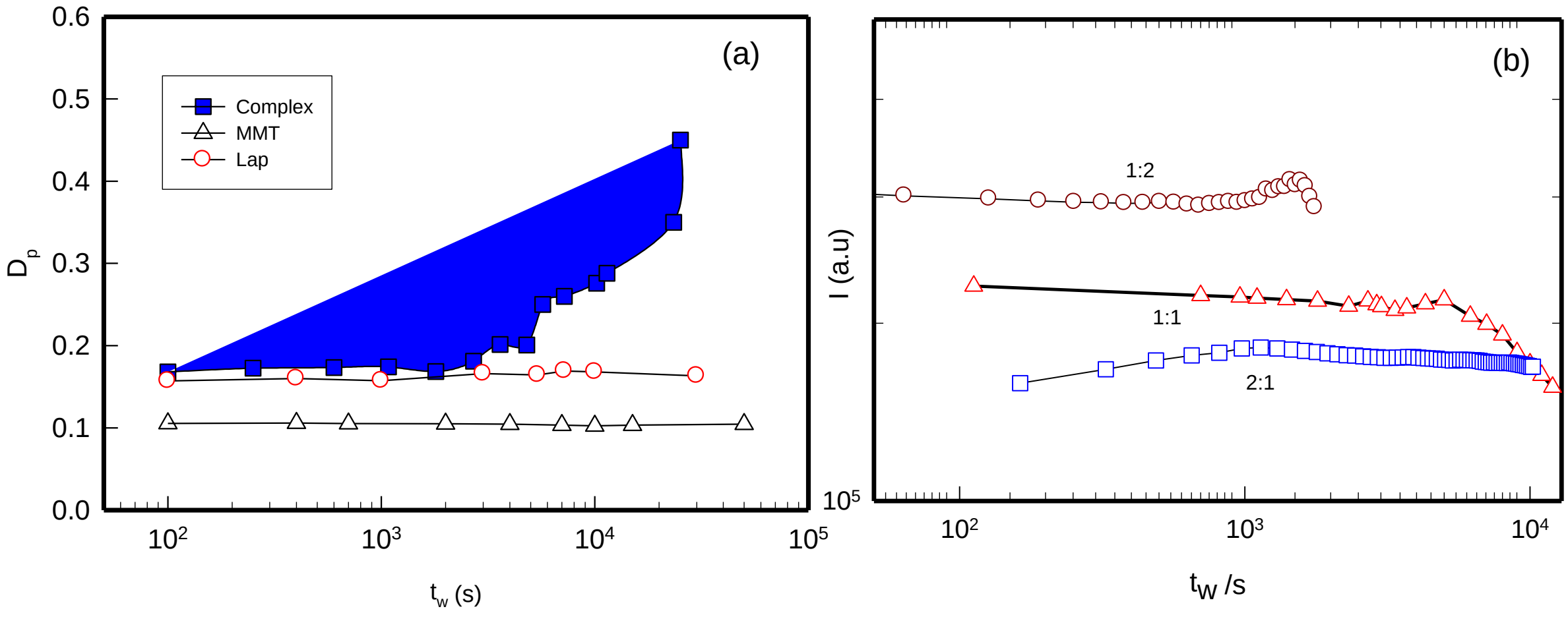

**Fig. S2** Left panel is depolarization ratio for Laponite, MMT and complex as a function of waiting time. The right panel is the growth of intensity with $t_w$.

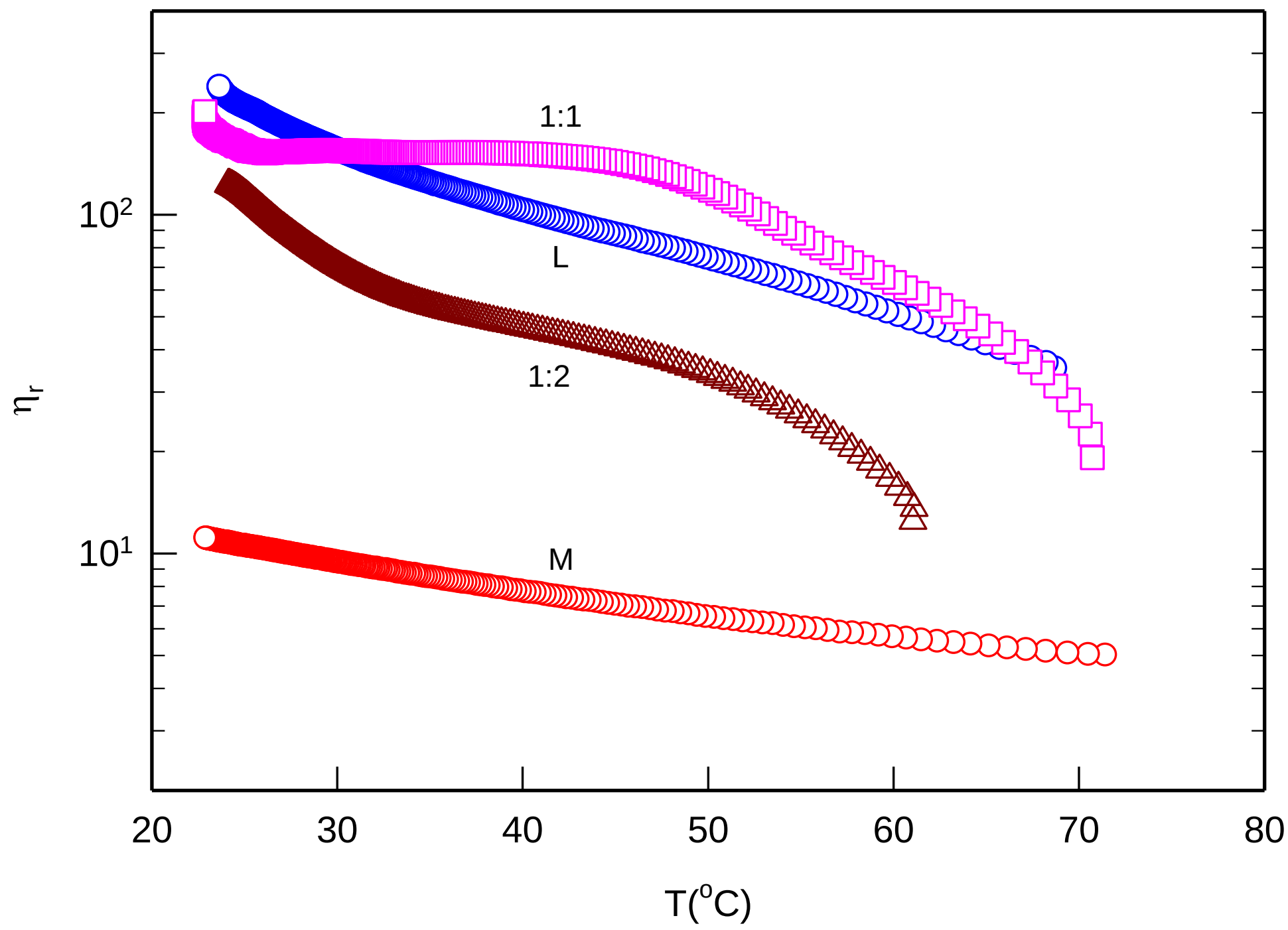

**Fig. S3** Cooling curves for the gels (heated up to 70°C) of different compositions as indicated. Single component curves yield more ordered cooling compared to the cogels.

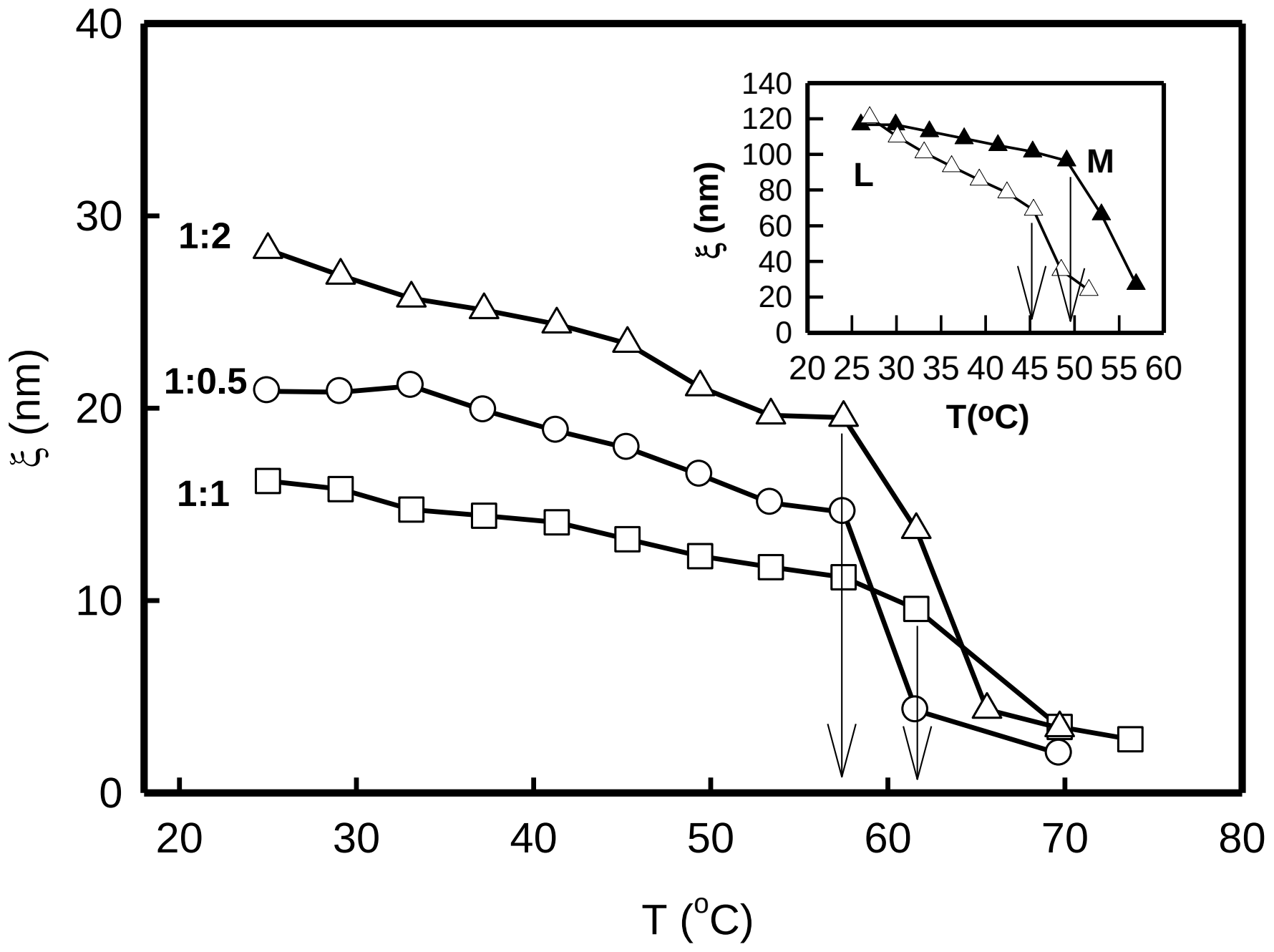

**Fig. S4** viscoelastic length as a function of temperature.